\documentclass[runningheads]{llncs}

\usepackage{graphicx}
\usepackage{comment}
\usepackage{todonotes}
\usepackage{tabularx}
\usepackage{url}
\makeatletter
\def\blfootnote{\gdef\@thefnmark{}\@footnotetext}
\makeatother

\begin{document}
\title{Dutch General Public Reaction on Governmental COVID-19 Measures and Announcements in Twitter Data}
\titlerunning{PuReGoMe Report}
\author{Shihan Wang\inst{1} \and
Marijn Schraagen\inst{1}\and
Erik Tjong Kim Sang\inst{2} \and
Mehdi Dastani\inst{1}}

\authorrunning{Wang, Schraagen, Tjong Kim Sang, Dastani}
\institute{Intelligent Systems, Utrecht University, The Netherlands \\
\email{s.wang2, m.p.schraagen, m.m.dastani@uu.nl}\\
\and
Netherlands eScience Center\\
\email{e.tjongkimsang@esciencecenter.nl}}
\maketitle  
\begin{abstract}
Public sentiment (the opinions, attitudes or feelings expressed by the public) is a factor of interest for government, as it directly influences the implementation of policies. Given the unprecedented nature of the COVID-19 crisis, having an up-to-date representation of public sentiment on governmental measures and announcements is crucial. While the ‘staying-at-home’ policy makes face-to-face interactions and interviews challenging, analysing real-time Twitter data that reflects public opinion toward policy measures is a cost-effective way to access public sentiment. In this context, we collect streaming data using the Twitter API starting from the COVID-19 outbreak in the Netherlands in February 2020, and track Dutch general public reactions on governmental measures and announcements. We provide temporal analysis of tweet frequency and public sentiment over the past nice months. We also identify public attitudes towards two Dutch policies in case studies: one regarding social distancing and one regarding wearing face masks. By presenting those preliminary results, we aim to provide visibility into the social media discussions around COVID-19 to the general public, scientists and policy makers. The data collection and analysis will be updated and expanded over time. 

\end{abstract}
\section{Introduction}
\blfootnote{This preprint was last updated on December 21, 2020.}

Public support is essential for the success of policy measures. Support can be measured from physical behaviours, but also from how people think and talk about these measures, which is known as public sentiment \cite{bakshi2016opinion}. As public sentiment (the opinions, attitudes or feelings expressed by the public) can directly influence the implementation of policies \cite{burstein2003impact}, it is crucial for policy makers to know the public sentiment of chosen policies and to take this sentiment into account when deciding on new policies. 

Given the unprecedented nature of the COVID-19 crisis, having an up-to-date representation of Dutch public sentiment on governmental measures and announcements becomes even more important. However, it is uncertain how public sentiment evolves when the perceived urgency of policy measures changes over time, which is our main research focus in this paper.

The ‘staying-at-home’ policy makes analysing public sentiment towards specific policies by means of face-to-face research methods like interviews and questionnaires challenging, while classical online surveys could delay the analysis results by the restriction of frequency.
Meanwhile, about 2.8 million users in the Netherlands use Twitter to share their opinions \cite{oosterveer2020}, making it a valuable platform for tracking and analysing public sentiment. It also allows for much more and frequent measurements and better indication of changes over time in public reactions. Therefore, aiming at understanding the variation of Dutch public sentiment during the COVID-19 outbreak period, we propose to analyse Twitter data using machine learning and natural language processing approaches \cite{tan2013interpreting,wang2015detecting}. By analysing real-time data through non-invasive methods, we aim to provide a cost-effective way to access public sentiment towards policy measures in a timely manner. 

In this paper we present the preliminary results of our data analysis. First, we collected Dutch Twitter data and filtered the data based on predefined COVID-19 related keywords. Second, we analysed the temporal pattern of Dutch public sentiment during the period of COVID-19 epidemic in the Netherlands. Third, we conducted two case studies to extract the public attitude towards governmental policies of social distancing and wearing face masks. 
In addition, to enhance the representative of the Dutch public opinion learned from Twitter data, we collect additional data from comments on the news website Nu.nl and Dutch posts on Reddit. This data is used for annotation, model training and analysis (similar to the Twitter data). 
By presenting those findings, we expect to provide a sentiment-oriented overview of social media discussions around COVID-19 to the general public, scientists and policy makers. 

\section{Related work}

Social media sentiment has previously been analyzed during pandemics such as H1N1 \cite{chew2010h1n1}.
Also the COVID-19 pandemic, despite being a relatively new topic, has attracted many researchers from different areas, including social media analysis. Abd-Alrazaq et al. \cite{abdalrazaq2020} noted four main COVID-19 related themes on Twitter: virus origin, contamination sources, preventive measures and impact on societies. Zhao et al. \cite{zhao2020descriptive} identified topics and sentiment related to COVID-19 in China. Samuel et al. \cite{samuel2020covid} conducts textual analyses of Twitter COVID-19 data to identify public fear sentiment. Several studies focus on bots \cite{ferrara2020} and misinformation \cite{kouzy2020,singh2020} which influence opinions on social media. Chen et al. \cite{chen2020government} show that positive polarity government messages on social media result in higher public engagement. Cinelli et al. \cite{cinelli2020various} identified COVID-19 related topics in various social media, including Reddit and Twitter. They found that the ratio of misinformation to reliable information was stable over time, however Twitter has a larger percentage of misinformation compared to Reddit.

From a technical point of view, our sentiment analysis covers two perspectives: polarity analysis (whether a message is positive or negative \cite{bakshi2016opinion}) and stance analysis (whether a message is supportive of or against a given target \cite{li2019multi}). Stance detection has recently received considerable attention in the NLP community \cite{kucuk2020stance}, for which neural networks with word embeddings have proven to be effective \cite{chen2017stance,li2019multi}. An annotation and training approach for classifying hate speech in COVID-19 related tweets using embeddings is described by Cotik et al. \cite{cotik2020hatespeech}.
Other approaches of interest for sentiment analysis include using lexical and semantic knowledge bases \cite{dridi2019semantics}, and unsupervised stance clustering methods \cite{darwish2020unsupervised}.

In more recent work, Miao et al. \cite{miao-etal-2020-twitter} studied the effect of the public's response on the pandemic measures in the United States by performing social media analysis. They found that data distillation of data sets of other domains was a good alternative for missing relevant training data. In related work, Addawood et al. \cite{addawood-etal-2020-tracking} showed that over 2020 the polarity of the public towards the pandemic degraded in Saudi Arabia while Kurten et al. \cite{kurten2020} performed a similar study for Belgium and Xue et al. \cite{xue2020} did a world-wide study. In other work based on tweets, Van Loon et al. \cite{loon-etal-2020-explaining} explored the relation between political preferences and tendency to socially distance.

%



\section{Data Acquisition and Processing}

Our main data set consists of Dutch tweets collected by the twiqs.nl service \cite{tjongkimsang2013dealing}. The analysis of this data is compared with comments from the discussion website Reddit and the Dutch news website Nu.nl, where the general public can add comments to news articles. We include data from February to November 2020 in this paper. 

\begin{table}[tp]
\begin{center}
    \begin{tabular}{lccc}
    \hline
\textbf{Month} & \textbf{Number of tweets}   & \textbf{Per day} & \textbf{Per hour}\\\hline
February 2020  & 14,851,678 & 512,126 & 21,338\\
March 2020     & 21,180,942 & 683,256 & 28,507 \\
April 2020     & 18,715,900 & 623,863 & 25,994 \\
May 2020       & 18,044,679 & 582,086 & 24,253 \\
June 2020      & 20,807,966 & 693,598 & 28,899 \\
July 2020      & 19,154,442 & 617,885 & 25,745 \\
August 2020    & 20,314,042 & 655,291 & 27,303 \\
September 2020 & 20,340,753 & 678,025 & 28,251 \\
October 2020   & 21,987,100 & 709,261 & 29,513 \\
November 2020  & 19,370,135 & 654,671 & 26,902 \\
    \hline
    \end{tabular}
    \caption{Total number of Dutch tweets collected in the study.}
    \label{tab-corpus-size}
\end{center}
\end{table}

\subsection{Twitter Data Collection}

Twiqs.nl is a service from the Netherlands eScience Center and Surf which collects Dutch tweets and provided aggregated analysis to the research community \cite{tjongkimsang2013dealing}. The service has been available since 2013 and currently harvests about 600,000 Dutch tweets per day. We use the data from February 2020 (the first month a COVID-19 patient was diagnosed in The Netherlands) onwards. We rely on the \textit{lang} feature provided by the Twitter API to determine the language that the tweet is written in. Tests with tweet replies indicate that our corpus covers more than 80\% of the tweets written in Dutch. An overview of the corpus size can be found in Table \ref{tab-corpus-size}. 

\subsection{COVID-19 Keyword Filtering}
\label{sec-filtering}

We filtered the general corpus to obtain only tweets that contain disease keywords, related topic words or related hashtags. The list of filter keywords is shown in Table~\ref{tab-filter-words}, which we selected based on four important categories related to our research focus. Keyword filtering was done case-insensitive and longer words containing one of these keywords as substrings, like the word \textit{coronavirus}, were also selected. The size of the filtered tweet corpus can be found in Table \ref{tab-selected-corpus-size}.

\begin{table}[tp]
\begin{center}
    \begin{tabular}{lll}
    \hline
    \textbf{Category} & \textbf{Keyword} & \textbf{English Translation}\\
    \hline
    Disease  & corona & \\
             & covid & \\
    \hline
    Health care  & huisarts & doctor\\
                 & mondkapje & face mask\\
    \hline
    Government & rivm & national health organization\\
    \hline
    Social     & flattenthecurve & \\
               & blijfthuis & stay home\\
               & houvol & hang in there\\
    \hline
    \end{tabular}
    \caption{Keywords used for filtering COVID-19 related tweets.} 
    \label{tab-filter-words}
\end{center}
\end{table}

The differences between the numbers of tweets in the various months are striking. We explore the reason of this diversity in Section~\ref{frequency}. We also observe some topic drift among selected tweets, for example in February the discussion was often about foreign outbreaks while in March that shifted to the measures (\textit{maatregelen}) taken by the Dutch government against the pandemic. Appendix~\ref{app-tweet-topics} contains an overview of all detected new topics in Dutch pandemic tweets.  

\subsection{Validation Data Collection: Reddit and Nu.nl}
\label{sec-reddit-nunl}

Considering the recent concern about groups amplifying minority opinions on Twitter \cite{diresta2018}, we analyse Dutch public opinion from two alternative data sets to validate our Twitter analysis results. In Table \ref{tab-selected-corpus-size}, we present the information of corpora from those data sources, Reddit and Nu.nl.

\begin{table}[tp]
\begin{center}
    \begin{tabular}{lrrr}
    \hline
\textbf{Month} & \textbf{Tweets}  & \textbf{Nu.nl} & \textbf{Reddit} \\\hline
February 2020  &   278,082 &  25,721 &      5 \\
March 2020     & 3,152,638 & 207,957 & 28,038 \\
April 2020     & 2,115,728 & 193,530 & 11,943 \\
May 2020       & 1,264,650 & 146,832 &  6,452 \\
June 2020      &   921,481 &  90,698 &  3,218 \\
July 2020      &   922,992 &  85,085 &  2,985 \\
August 2020    & 1,078,644 & 105,047 &  4,738 \\
September 2020 & 1,115,057 & 114,533 &  6,486 \\
October 2020   & 1,571,318 & 193,579 & 11,013 \\
November 2020  &   809,093 & 113,924 &  4,840 \\
    \hline
    \end{tabular}
    \caption{Size of on-topic corpora used for our analysis, measured in total number of tweets, total number of Nu.nl comments and total number of Reddit posts per month.}
    \label{tab-selected-corpus-size}
\end{center}
\end{table}

Nu.nl is a widely used Dutch online news website. This website publishes news articles 24 hours a day, and allows the public to respond to every individual news article by posting comments. We have collected all comments on news articles tagged \textit{Coronavirus}\footnote{\url{https://www.nu.nl/tag/Coronavirus}}.

Reddit is organized in a large number of high-level topics called \textit{subreddits}. Users can post a message in a subreddit in order to start a \textit{thread}, and other users can post comments in reply to either the original message or an existing comment. There are a large number of threads related to COVID-19 on Reddit, 
for instance a thread called ``Megathread Coronavirus COVID-19 in Nederland'' in the subreddit r/thenetherlands has been active since March 2020. This thread was one of the main sources of the Reddit posts included in this comparison analysis. 
We collected all comments in the subreddits \textit{coronanetherlands}, \textit{CoronaNL}, \textit{CoronavirusNL} and the \textit{Megathread Coronavirus COVID-19 in Nederland} of the subreddit \textit{thenetherlands}. As we considered only the Dutch comments in this study, we excluded about 10\% of English comments in the Reddit data using the Python package langid \cite{lui2012}.

All collected Nu.nl and Reddit comments are assumed to be on-topic, therefore no keyword filtering has been applied during data collection. Note that the Nu.nl and Reddit data sizes are orders of magnitude smaller than the tweet corpus
(see Table \ref{tab-selected-corpus-size}).

\begin{figure}[t]
\includegraphics[width=\textwidth]{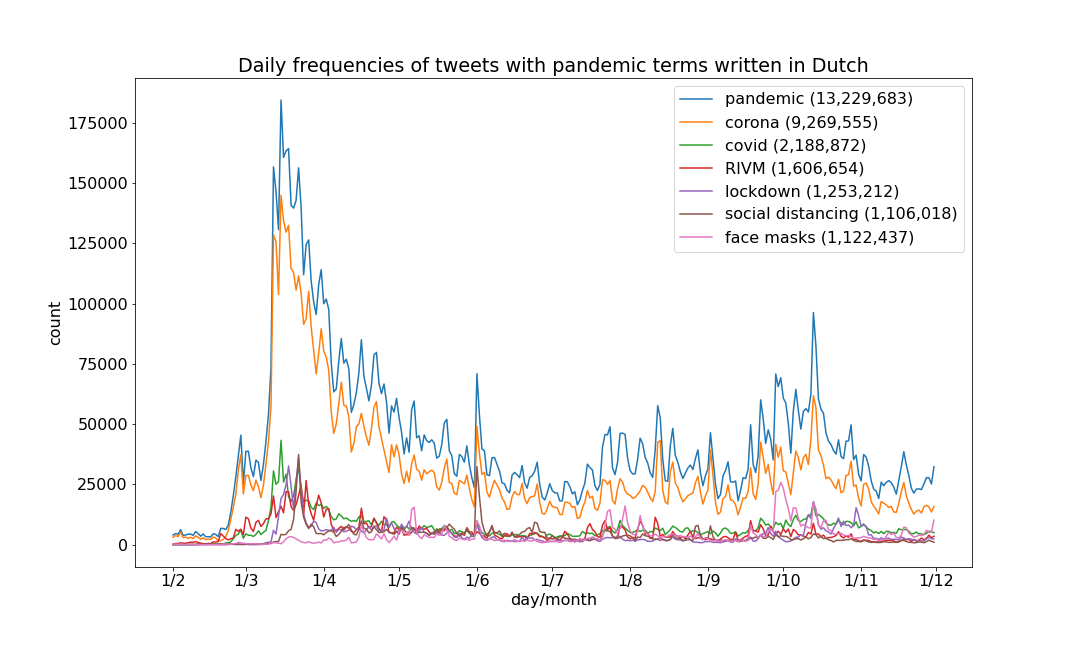}
\caption{Daily frequencies of Dutch tweets related to the COVID-19 topics. The line of `pandemic' covers tweets with any of the keywords in Table \ref{tab-filter-words}, while the other lines represent individual keywords. RIVM is the Dutch national health institute. The peak of each line is around the date of an important pandemic press conference by the Dutch government (March 12th, 2020).}
\label{fig-tweet-frequencies}
\end{figure}

\section{General Public Reaction and Sentiment}

We perform a temporal analysis of Dutch public reaction by examining the frequency of the COVID-19 related tweets and average sentiment (polarity) of those tweets. We also take the regular press conference of the Dutch government into account to analyse the correlation between public reaction and government policies over the time period.

\subsection{Temporal Frequency of COVID-19 Related Tweets}
\label{frequency}

The daily frequencies of Dutch COVID-19 tweets are presented in Figure \ref{fig-tweet-frequencies}. We define COVID-19 related tweets as messages written in Dutch that contain any of the eight keywords listed in Table~\ref{tab-filter-words}
(case-insensitive and possibly as substring). We found the most popular five keywords (i.e. \textit{corona}, \textit{covid}, \textit{rivm}, \textit{lockdown} and \textit{mondkapje}) and also plotted frequencies of tweets containing one of these keywords. Finally, the frequency of tweets on social distancing was added (see section \ref{sec-soc-dist} for a definition of such tweets).

In Figure \ref{fig-tweet-frequencies}, the COVID-19-related tweets peak around a press conference of the Dutch government on Thursday March 12th 2020 (in which the government announced the first national lockdown measures to stop spread of corona virus in The Netherlands) and reach the top after March 15th 2020 (when the bars and restaurants were closed). After that, the COVID-19-related tweets keep decreasing in frequency. As noted in Table~\ref{tab-selected-corpus-size}, the number of tweets containing the selected keywords decreases over time until July when it started rising again. There are two possible reasons for the decrease, either the topic became less popular or people continue to talk about COVID-19 in tweets but without using the keywords in Table~\ref{tab-filter-words} such as \textit{corona} or \textit{covid}.

\begin{figure}[tp]
\begin{minipage}[t]{0.48\textwidth}
\begin{center}
\includegraphics[width=1\textwidth]{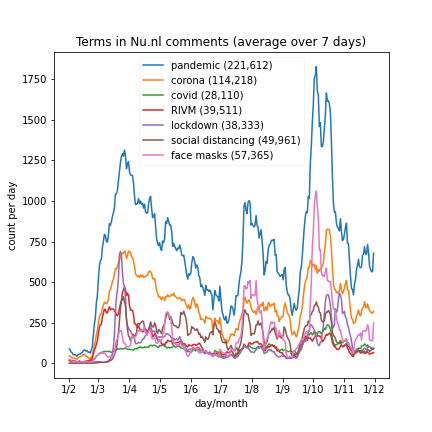}
\end{center}
\end{minipage}
\begin{minipage}[t]{0.48\textwidth}
\begin{center}
\includegraphics[width=1\textwidth]{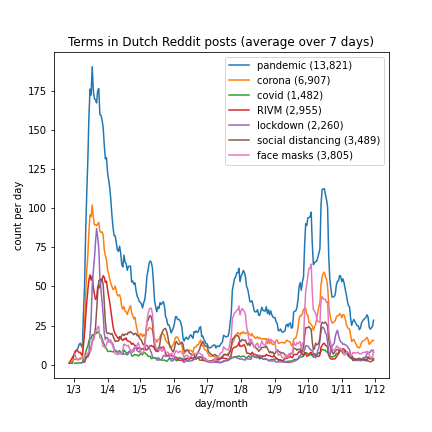}
\end{center}
\end{minipage}
\caption{Daily frequencies of some pandemic terms in Nu.nl comments (left) and Dutch Reddit posts (right). The lines of `pandemic' covers tweets with any of the keywords in Table \ref{tab-filter-words}. RIVM is the Dutch national health institute.}
\label{fig-reddit-keywords}
\end{figure}

To check these hypotheses a similar analysis is performed on the data sets of Reddit and Nu.nl and their results are presented in Figure~\ref{fig-reddit-keywords}. The left sub-graph of Figure~\ref{fig-reddit-keywords} shows the frequency of Reddit comments per day related to COVID-19, which is consistent with the Twitter data. In this case the decrease may be caused either by people finding other threads to post comments, or a general loss of interest in the topic. Further analysis shows that the relative frequency of keywords in the collected threads remains rather stable. Because all the messages are about COVID-19 by default, the relative frequencies are indicative of the way people talk about the topic. This data combined with the tweet analysis therefore seems to support the hypothesis that the topic was actually less popular in July 2020 compared to the preceding period, and the hypothesis that people use different words to discuss the topic is not supported. 

Interestingly, analysis also shows that a large number of the messages in the Reddit and Nu.nl comments did not use any of the four popular keywords in Figure \ref{fig-tweet-frequencies}, while all messages are about COVID-19. One explanation for this may be that within long comment threads the messages may become less self-contained than, e.g., a new tweet, which could influence the usage of keywords. However, it is also possible that the Twitter analysis suffers from a large false negative rate of COVID-19 related tweets that are not selected by the keyword filter, even in the peak period around March 15th where keyword frequencies are the highest.

\begin{figure}[t]
\includegraphics[width=\textwidth]{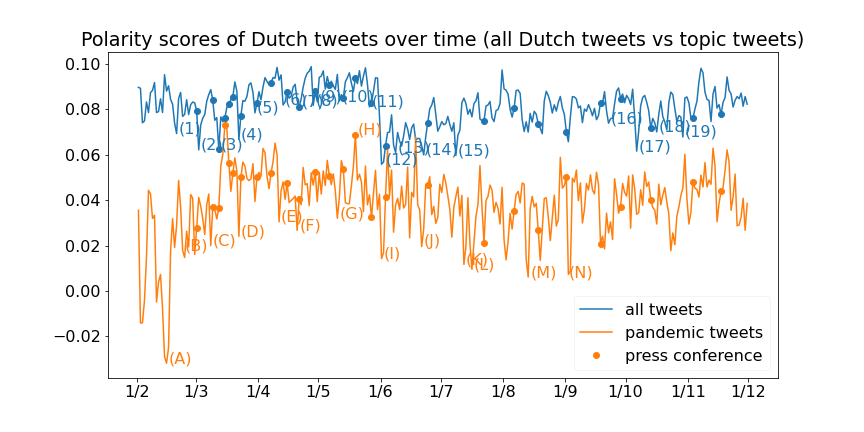}
\caption{Average daily sentiment of all Dutch tweets (blue line) in comparison with average daily sentiment of the COVID-19 topic related tweets (orange line). The dots indicate dates of government press conference about the pandemic, whose exact date and detailed contents are given in appendix \ref{app-press-conference-topics}. We also listed the content of several events that happened on the date of local minima of the two lines in appendix \ref{app-local-minima-topics}.} 
\label{fig-sentiment-topic-vs-all}
\end{figure}

\subsection{Temporal Sentiment of COVID-19 Related Tweets}

We used the sentiment module of the Python library \texttt{pattern} \cite{desmedt2012} to automatically assign a polarity score to each tweet. The scores are based on \texttt{pattern}'s Dutch sentiment lexicon which contains 3918 words, mostly adjectives, and a language-independent emoji lexicon with about 120 emoji. The sentiment scores range between -1 (very negative) and +1 (very positive). Two-thirds of the tweets received a non-zero score. An overview of the daily average sentiment can be found in Figure \ref{fig-sentiment-topic-vs-all}. To assess the results, we compare our daily tweet polarity scores with those of a similar study by the Dutch national statistics office \cite{cbs2020} and found a positive correlation (r=0.46 for February-June). 

In Figure \ref{fig-sentiment-topic-vs-all}, the y-axis represents the polarity score calculated by our sentiment analysis approach, where a higher polarity score means more positive attitude. We can observe that the average polarity score of Dutch tweets related to COVID-19 was always lower than the general polarity score, which indicates that, compared with general topics discussed in Twitter, the public is more negative towards COVID-19-related topics. Also, some interesting links can be found between press conferences and trends of public sentiment. For instance, general sentiment reached a low peak point on the date of the third press conference of the Dutch government about the pandemic, March 12th (3), when the first lockdown measures were announced. More recently, the COVID-19-related sentiment reached a high peak point on the date of May 19th (I), when the government announced the first release measures. Because of these findings we conjecture that the trend of average daily sentiment shown in Twitter is influenced by governmental measures and announcements.

\begin{figure}[t]
\begin{center}
\includegraphics[scale=0.33]{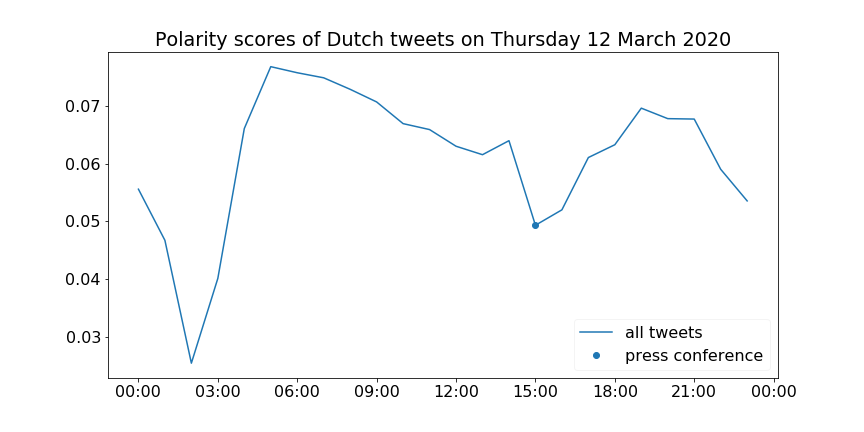}
\end{center}
\caption{Average daily sentiment of all Dutch tweets on the date of the third Dutch pandemic press conference about the first lockdown measures (March 12th, 2020).} 
\label{fig-sentiment-20200312}
\end{figure}

In addition, we studied the temporal sentiment development for all Dutch tweets on March 12th in detail to investigate the short-term influence of governmental announcements. 
We found that the impact of the press conference is visible in the evolution of the overall sentiment measured per hour. As shown in Figure \ref{fig-sentiment-20200312}, the sentiment is relatively stable from 05:00h onwards, but suddenly drops around the time of the press conference at 15:00h. The result indicates that the reaction of public sentiment towards governmental measures and announcements can be captured in a very short time period from Twitter data. 

\section{Case Studies: Public Attitude towards Specific Policies}

Besides the above application of sentiment analysis for COVID-19 in general, we are also interested in public attitude towards certain topics (e.g. a specific policy measure). In particular, we study whether the public opinion is supportive of or against a specific policy measure, which is known as stance analysis \cite{kucuk2020stance}. In this paper, we analyse the stance with respect to two national pandemic measures: social distancing and wearing face masks.

We employed the machine learning library fastText for determining the stance of tweets automatically. This library is used for on-the-fly training of word embeddings with subword features from the input data, as part of a linear feed-forward neural network \cite{bojanowski2017enriching,joulin2017bag}.
For each topic a classifier is trained with manually annotated training data. After evaluating these models, the classifiers are used to predict the stance of all other messages in the dataset in order to perform a comprehensive stance analysis for the target topics.

\subsection{Case study 1: social distancing}
\label{sec-soc-dist}

In our first case study, we assessed the stance of Dutch Twitter with respect to \textit{social distancing}. This measure was announced nationally on Sunday 15 March 2020. It recommends the public to keep one-and-a-half meters distance to each other during social interactions. Note that this definition (and the corresponding data collection) excludes the closely related topic of avoiding social interaction with other people.

\subsubsection{Data processing and annotation}
We built a query for social distancing tweets interactively. Starting with the Dutch word for one-and-a-half (\textit{anderhalve}), we compared the tokens in matched and unmatched Dutch tweets of 1 June 2020 with the t-score \cite{church91}. This produced three more interesting query words: 1.5, meter and \textit{afstand} (Dutch for distance). We repeated the process in an iterative manner until no useful new query words appeared. In the end, 
we use the query \textit{anderhalve} followed by \textit{meter}, or \textit{1.5} or \textit{1,5} followed by \textit{m}, or both \textit{afstand} and \textit{hou*} (Dutch for ``to keep'') anywhere in the tweet in any order.\footnote{The following regular expression was used for implementation:\\\verb~1[.,]5[ -]*m|afstand.*hou|hou.*afstand|anderhalve[ -]*meter~} 

\begin{figure}[t]
    \centering
    \includegraphics[scale=0.3]{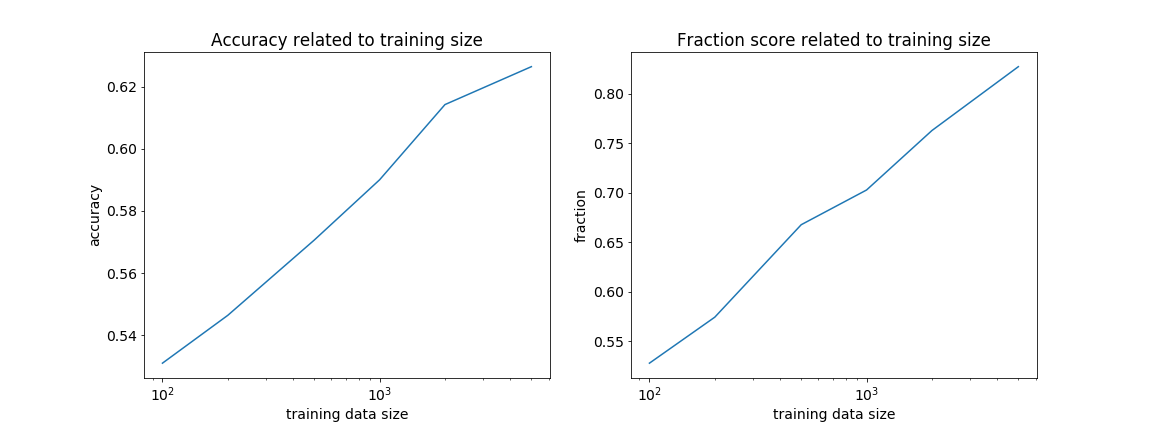}
    \caption{Size of the training data related to the labeling performance of the best model of the machine learner fastText measured on held-out test data. The larger the training data, the better the performance of the learner. There is no indication of the performance plateauing and optimal performance seems to require more than 100,000 (10$^5$) manually labelled tweets. However we regard a 0.8 fraction score as quite satisfactory.}
    \label{fig-training-size}
\end{figure}

We applied the defined query to our Dutch tweet collection of February-June 2020 and found 670,249 matching tweets (including retweets). We randomly selected about 1\% of the tweets (5,732 unique tweets) for annotation. These tweets were labelled by a single annotator with respect to the question: \textit{Does the tweet or tweet author support or reject the social distancing measure announced by the Dutch government on 15 March 2020?}. The annotation scheme included three labels: Supports, Rejects and Other (for irrelevant tweets and tweets for which the stance could not be determined). The annotator found 56\% Supports tweets, 24\% Rejects tweets and 20\% Other tweets. 400 tweets were annotated by a second annotator to access the inter-annotator agreement, which proved to be $\kappa=0.60$ \cite{cohen1960}. 

\subsubsection{Stance classifier training and evaluation}
For training the fastText model for stance analysis, we used ten-fold cross-validation on the human labeled tweets in combination with a grid search to determine the optimal word vector dimension (10-300), number of training epochs (10-500) and learning rate (0.05-1.0). FastText consists of an integrated neural network architecture which estimates useful word representations with a skipgram model \cite{bojanowski2017enriching} and classifies texts with three-layer network based on a linear model \cite{joulin2017bag}. For the setup of the neural network itself we used the default settings of the fastText library. The dataset was divided in 80\% training, 10\% validation for parameter optimization, and 10\% test examples. 

For our application, accurately predicting the stance label of each text in the data is very expensive and unnecessary, given that we care more about the public attitude rather than individual opinion. We therefore introduce the \textit{fraction score}, which measures the performance of predicting the relation between \textit{Reject} labels and \textit{Support} labels. The effectiveness of such proportion measurement has been proved within practical prediction problems \cite{forman2005,card2018}. 
In this paper, we define the fraction score as $r_{predicted}/r_{gold}$, where $r$ is:
\begin{equation}
    r = \frac{Number_{Reject~labels}}{Number_{Support~labels}}
\end{equation}

We optimize the parameter settings based on both the accuracy of individual prediction and the performance of fraction score. The optimal value of the fraction score is 1.0, however given that the classifiers predict more \textit{Support} than \textit{Reject} labels their fraction scores are lower than one. The optimal parameter settings and classification results are shown in Table \ref{tab:results}. The baseline method labels the given message as the majority class (i.e., \textit{support} in case of social distancing tweets). 

\begin{table}[t]
    \centering
    \begin{tabular}{lr|r}
    \hline
         &  Evaluated by accuracy & Evaluated by fraction score\\
         \hline
    vector length     & 10 & 200 \\
    learning rate     & 0.2 & 0.2\\
    epochs            & 10 & 200\\
    \hline
    baseline performance        & 0.56 & 0.00\\
    validation performance & 0.65 & 0.88\\
    test performance     & 0.65 & 0.83\\
    \hline
    \end{tabular}
    \caption{The parameter setting and performance of the trained stance classifier for social distancing tweets}
    \label{tab:results}
\end{table}

To assess the effect of the training size on the performance of the automatic classifier, we performed additional 10-fold cross-validation experiments with varying training sizes (but constant test sizes). The results can be found in Figure \ref{fig-training-size}. Both accuracy and fraction score increase for larger training sizes, while there is no indication of the performance lines leveling off. This indicates a potential for performance improvement with more annotated training data. 


\subsubsection{Stance analysis}
\begin{figure}[t]
    \centering
    \includegraphics[scale=0.43]{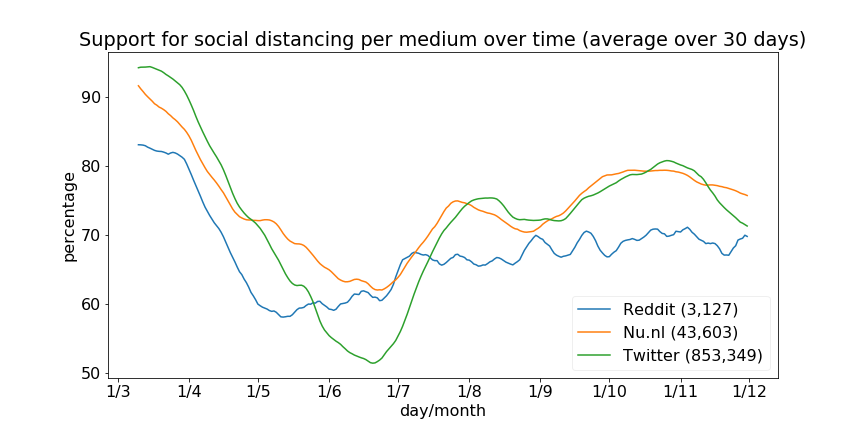}
    \caption{The variation of public support for the March policy on social distancing on Reddit (blue), the Dutch news site Nu.nl (orange) and Twitter (green). After initial large support for the measure in March, the support drops to the lowest levels in May/June, when the number of infections was also low (see Figure \ref{fig-corona-numbers}). Since then the support has increased.}
    \label{fig-social-distancing}
\end{figure}

Next, we applied the trained fastText model (optimized by the fraction score) for labelling the tweets of February--November 2020. The temporal rates of \textit{Supports}, \textit{Rejects} and \textit{Other} tweets can be found in Figure \ref{fig-social-distancing} (green line). After initially being high in March (95\%), the support gradually decreases to below 50\% in June. This is consistent with reports by the Dutch health authority RIVM on public support for the measure \cite{rivm2020}. On the other hand, the graph also shows an increasing support on Twitter for the social distancing measure in June and July, which was not immediately reported by the health authorities. Eventually the rise in support for the measure was confirmed by the RIVM in September 2020 \cite{rivm2020b}, based on questionnaires filled in by its general public panel. The fact that social media analysis picked up this change supports the validity of this type of analysis. 


A confounding issue is the activity of fringe groups amplifying minority opinions on Twitter \cite{diresta2018}, which may influence stance measurements based on tweets. For this reason we applied the same classification model to assess the stances of messages on social distancing from the other two platforms, Reddit and Nu.nl. The temporal stances on all three platforms are presented in Figure \ref{fig-social-distancing}. We can find a similar trend among all three platforms that validates our Twitter analysis. The related data volume on Reddit and Nu.nl is a lot lower than on Twitter. For Reddit, as only around 3000 messages are found in 5 months, the result may have been influenced by noise. 

\subsection{Case study 2: face masks}

In our second case study, we focused on the specific Dutch policy measure on wearing face masks (Dutch: \textit{mondkapjes}). At the start of the pandemic the Dutch health organization RIVM advised against the use of face masks by the general public because it deemed the other measures (social distancing in combination with frequent hand washing) sufficient. Since June 2020 the policy has changed gradually, with face masks being mandatory in public buildings since December 2020. For annotation we used the original policy, for which the annotator answered the question \textit{What is the opinion of the tweet about the RIVM policy advising against the use of face masks by the general public?}. Similar to the social distancing data, 578 tweets and 744 Nu.nl comments that contained the word {\textit mondkapje} were annotated by a single annotator with one of three labels: \textit{Supports}, \textit{Rejects} and \textit{Other}. Then, a trained fastText model was built from the labelled data. Based on this model, the labelling results of all tweets, comments and posts containing the word \textit{mondkapje} can be found in Figure \ref{fig-stance-mondkapje}.

We found that most of the tweets (85\%) and most Nu.nl comments (75\%) rejected the policy. There were too few Reddit posts on this topic to obtain an accurate picture over time but the stance was also less than 50\% supportive. Two examples of frequent opinions (translated to English) are as follows:

\begin{figure}[t]
    \centering
    \includegraphics[scale=0.43]{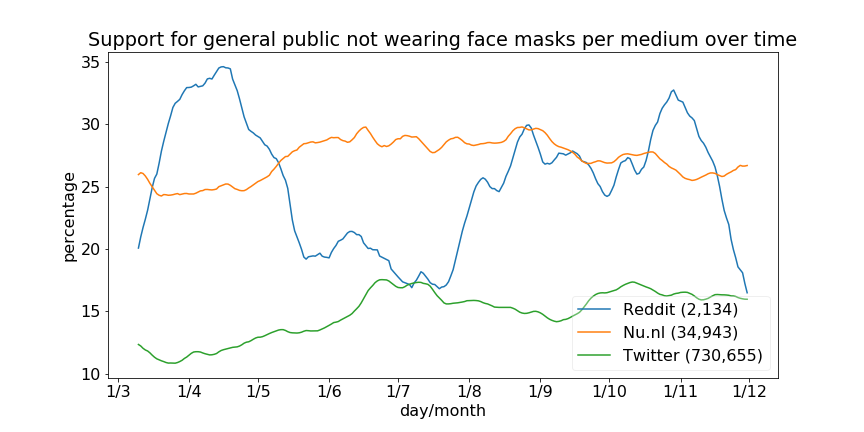}
    \caption{Development of the support for the April policy on the general public not wearing face masks on Reddit (blue), the Dutch news site Nu.nl (orange) and Twitter (green). On Twitter and on Nu.nl there is a rejection of the policy over time at stable rates, although at different absolute levels. On Reddit the policy is rejected as well but the rate shows more variance over time. Note that the Reddit graph was based on fewer data (2,134 posts vs 34,943 Nu.nl comments and 730,655 tweets). Face masks became mandatory in Dutch public transport in June and in Dutch public buildings in December.}
    \label{fig-stance-mondkapje}
\end{figure}

{\renewcommand{\arraystretch}{1.5}

\noindent\begin{tabular}{p{\textwidth}}
\hline
\textit{Face masks are useless: corona droplets do not travel more than 1.5m. Is this a scientific conclusion of the RIVM? Or is it a fake argument because of shortages? 2008 RIVM scientist says that face masks reduce infections and have effect.}\\
\hline
\end{tabular}

\noindent\begin{tabular}{p{\textwidth}}
\hline
\textit{Broadly I agree. But with respect to face masks I refer to scientists in our neighboring countries. They have another opinion than the RIVM, So yes$...$ who is right?}\\
\hline\end{tabular}
}

A possible reason for the stance being more supportive on Nu.nl than on Twitter is that the comments on Nu.nl are actively moderated while tweets are not, leaving less room for trolls to attack the government policy. This finding also indicates the importance of capturing the public reactions across different sources. 

\section{Discussion and Future Work}

We present a first analysis of Dutch tweets related to the COVID-19 pandemic. We concentrate on online public reactions towards the measures taken by the government against the spread of the corona virus. Our analysis shows that the pandemic generated large interest on Dutch Twitter with tens of thousands of tweets on the topic each day
. The sentiment of the COVID-19-related topic tends to be more negative than that of Dutch tweets on average
. We also found that national press conferences about governmental measures and announcements (e.g. the first national press conference about the lockdown policy) can influence public sentiment, and furthermore the impact on tweet sentiment can be captured in a very short time period. 
However, we did not find a direct correlation between all press conferences and trends of public sentiment. We would like to investigate the influence of governmental measures and announcements in more detailed analysis. In future work, we also see the potential to explain temporal trends of the public sentiment by linking these trends to daily numbers of infections and deaths in the Netherlands.

In two case studies we assessed the stance of Dutch tweets on the national advice on social distancing and on wearing face masks. The analysis showed that people widely supported social distancing when the measure was announced in March, then support declined until June and increased again recently. We think this phenomenon may be related to the the pandemic situation in the Netherlands (the number of reported COVID-19 patients has also decreased and increased during this time, as shown in Appendix \ref{app-number}).
With respect to face masks, analysis showed that the public widely rejected the initial government position that face masks are useless for the general public. The rejection rates remained relatively stable over the past months.

As future work, we are interested in performing stance analysis regarding other policy measures, like compulsory school closing and infection testing. In order to be able to tackle stance across topics with machine learning, some initial manual annotation needs to be done for each of these new topics. Active learning \cite{dasgupta2011} could be an interesting approach for limiting the burden of this task. This method identifies the parts of the data for which annotation would be most beneficial to the machine learner and thus limits the effort required for performing the manual annotation process.  

From a practical perspective, we would like to improve the explainability of our analysis results. We are collaborating with social scientists, who use quantitative methods (i.e. online questionnaires and interviews) to investigate the influence of policy measures on the general public.
 
We are also interested in a geographic separation of the data. Some tweets contain location information which would be useful for detecting Dutch tweets from other countries, like Belgium, which we want to exclude. But the location information could also be used for dividing The Netherlands in regions and look for regional differences in attitudes towards government policies. If such differences were found, they could possibly help explain regional variances in the spread of the virus.

\bibliographystyle{splncs04}
\bibliography{bib}

\pagebreak

\appendix
\section{Daily pandemic numbers in the Netherlands}
\label{app-number}

\begin{figure}
    \centering
    \includegraphics[scale=0.33]{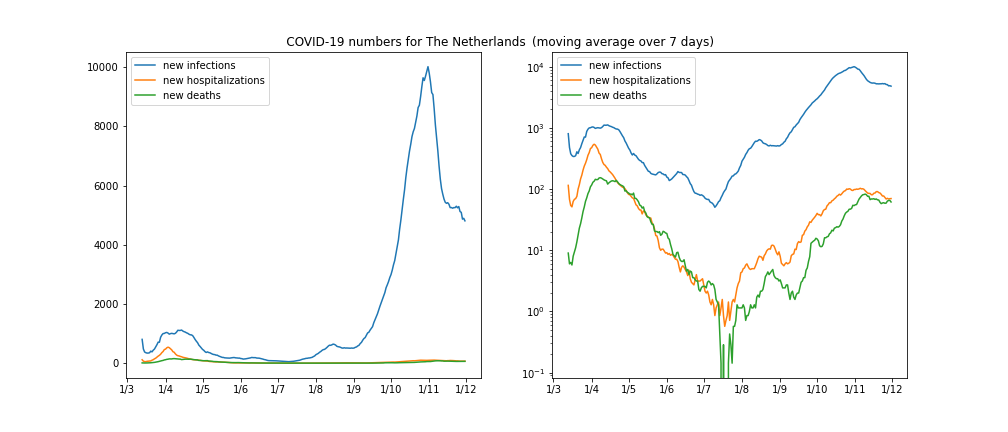}
    \caption{Daily number of infected people, hospital admissions and deaths in the Netherlands (moving average over 7 days). Note that during the first wave (March/April), there was insufficient test capacity so the reported numbers of detected new infections for this period are much lower than the real numbers.}
    \label{fig-corona-numbers}
\end{figure}

\vfill
\pagebreak

\section{Press conference topics}
\label{app-press-conference-topics}

\begin{tabularx}{\textwidth}{l|p{8.0cm}}
\textbf{Date} & \textbf{Measures}\\\hline
Sunday 1 March 2020
& People with respiratory issues should stay at home\\
Monday 9 March 2020
& Wash hands regularly. Sneeze in elbow. Use paper handkerchiefs. Stop shaking hands. Brabant only: work at home if possible. Spread working hours.\\
Thursday 12 March 2020
& No events with more than 100 people. Museums and theaters close. Sporting events cancelled. Work at home when possible. Vulnerable people avoid social contact. \\
Sunday 15 March 2020
& Schools close. Bars, restaurants and sport clubs close. Keep a distance of 1.5 metres from each other.\\
Tuesday 17 March 2020
& Government will provide financial aid for small and large corporations and for hospitals.\\
Thursday 19 March 2020
& No visits allowed in care homes.\\
Monday 23 March 2020
& No meetings of more than two people in the public space punishable by fines. Whole household self-quarantines if one member has a fever.\\
Tuesday 31 March 2020
& Measure deadline extended to 28 April (was 6 April). \\ 
Tuesday 7 April 2020
& Government thinks about a contact tracing app.\\
Wednesday 15 April 2020
& No new measures.\\ 
Tuesday 21 April 2020
& Primary schools partially reopen on May 11th. Children may practise sports outside from April 28. Events before September 1st are cancelled. \\ 
Wednesday 29 April 2020
& No new measures.\\ 
Wednesday 6 May 2020
& Avoid crowded places. Testing will be available to everybody from June 1st. Road map announced: 11 May: contact professions (for example barbers and driving instructors) may start working. Grownups may start sporting outside. Libraries reopen on. 1 June: Compulsory wearing of face masks in public transport. High schools reopen partially. Bars, restaurants, museums and cinemas reopen on: max 30 guests inside, no limit outside. 1 July: limit is raised to 100 people inside, also for churches and weddings. Public toilets reopen. 1 September: Sporting inside is allowed.\\ 
Wednesday 13 May 2020
& No new measures. \\
Tuesday 19 May 2020
& Primary schools reopen completely on 8 June. Partial reopening of universities on 15 June. There will be contact tracing of positive cases. Visitors allowed in care homes from 15 June.\\
Wednesday 27 May 2020
& All measures of 1 September move to 1 July. Spread holiday times. Only travel to corona-safe countries. \\\hline
\multicolumn{1}{l}{\small (continued on next page)} & \\
\end{tabularx}

\begin{tabularx}{\textwidth}{l|p{8.0cm}}
\textbf{Date} & \textbf{Measures}\\\hline
Wednesday 3 June 2020
& Travelling from and to most European countries possible from 15 June, negative advice for other countries. Primary schools reopen.\\
Wednesday 24 June 2020
& Larger gatherings of people allowed from 1 July. Secondary schools reopen.\\
Wednesday 22 July 2020
& No new measures or measure releases.\\
Thursday 6 August 2020 
& Limits on events for students. Bars must register all clients. \\
Tuesday 18 August 2020
& No more than six guests allowed at home. \\
Tuesday 1 September
& No new measures.\\
Friday 18 September
& Regional measures for badly affected urban areas: bars close at 00:00.\\
Monday 28 September
& Bars close at 22:00. Office workers should work at home. No audience at sport matches. Face masks advised when shopping in badly affected urban areas (became a national advise on Wednesday 30 September).\\
Tuesday 13 October
& Partial lock down: bars and restaurants close. Alcohol sale stops at 20:00. Maximum of 30 persons together inside. Maximum of 3 guests at home per day. No more team sports for adults. No more mass events. No shop opening in the evening.\\
Tuesday 3 November
& Partial lockdown remains until half December. Extra measures: advice to stay at home. Maximum group size outside: 2. Maximum of 2 guests at home per day. Museums and zoos close. Advice not to travel abroad.\\
Tuesday 17 November
& Extra measures of Tuesday 3 November end. Back to partial lockdown: bars and restaurants remain closed. Alcohol sale forbidden after 20:00. Maximum of 30 persons together inside. Maximum of 3 guests at home per day. No more team sports for adults. No more mass events. No shop opening in the evening\\
\hline
\end{tabularx}

\pagebreak
\section{Event topics} 
\label{app-local-minima-topics}

\begin{tabular}{ccl}\hline
\multicolumn{3}{l}{\textbf{Events in the blue line of Figure~\ref{fig-sentiment-topic-vs-all} (all tweets)}}\\\hline
 (1) & 20200220 & terrorist attack Hanau, Germany \\
 (2) & 20200302 & newsitem: Netherlands financially supporting asylum seekers \\
 (3) & 20200312 & announcement of first national COVID-19 measures \\
 (4) & 20200322 & lack of social distancing in nature areas \\
 (5) & 20200330 & news item: EU donates millions to Morocco \\
 (6) & 20200413 & news item: anti-gay violence \\
 (7) & 20200421 & national press conference on COVID-19 \\
 (8) & 20200428 & news item: lack of social distancing at IKEA \\
 (9) & 20200430 & news item: local coalition of CDA and FvD \\
(10) & 20200511 & news item: ethnic profiling by tax services \\
(11) & 20200526 & news item: KLM boss receives bonus \\
(12) & 20200601 & lack of social distancing at Amsterdam BLM demonstration \\
(13) & 20200608 & attacks on left-wing politicians \\
(14) & 20200622 & racism in tv programme Veronica Inside \\
(15) & 20200708 & farmers protest \\
(16) & 20200922 & Dutch VIPs involved in protest against measures (\#ikdoenietmeermee) \\
(17) & 20201006 & Eddie Van Halen dies \\
(18) & 20201016 & Paris teacher killed \\
(19) & 20201029 & Nice terror attack \\\hline
\multicolumn{3}{l}{\textbf{Events in the orange line of Figure~\ref{fig-sentiment-topic-vs-all} (pandemic tweets)}}\\\hline
(A) &20200215& COVID-19 death on Dutch cruise ship\\
(B) &20200223& Italy outbreak\\
(C) &20200308& news item: COVID-19 spreading in Europe\\
(D) &20200322& news item: lack of social distancing in nature areas\\
(E) &20200411& news item: Netherlands sent face masks to China\\
(F) &20200420& anti-climate protest on Twitter\\
(G) &20200510& warnings about staying alert\\
(H) &20200519& relaxation of COVID-19 measures announced\\
(I) &20200601&lack of social distancing at Amsterdam BLM demonstration\\
(J) &20200621& The Hague demonstration cancelled for lack of social distancing\\
(K) &20200712& news item: face mask related murder in France\\
(L) &20200716& discussion about initial health care worker safety\\
(M) &20200813& coalition leaves parliament to avoid vote on health care worker salaries\\
(N) &20200902& new photos showing lack of social distancing wedding Grapperhaus\\
\hline
\end{tabular}
\vfill

\section{Relevant new topics found in Dutch pandemic tweets}
\label{app-tweet-topics}

\begin{tabular}{lp{10cm}}
20200204 & First COVID-19 case in Belgium\\
20200205 & Japanese cruise ship infections\\
20200206 & Chinese doctor Li Wenliang dies\\
20200208 & Absence of testing at Dutch airport Schiphol\\
20200215 & COVID-19 patient on Dutch cruise ship Westerdam\\
20200217 & Dutch Red Cross starts collection for foreign COVID-19 aid\\
20200218 & Malaria medicine seems to help against COVID-19\\
20200222 & Incubation period could be longer than two weeks\\
20200223 & Rising number of infections in Italy\\
20200224 & Calls for more Dutch action on COVID-19\\
20200225 & Quarantine on Tenerife because of Italian COVID-19 patient\\
20200226 & Positive German case had recently visited The Netherlands\\
20200227 & First COVID-19 case in The Netherlands\\
20200228 & Second COVID-19 case in The Netherlands, in Amsterdam\\
20200229 & Seventh COVID-19 case in The Netherlands, in Delft\\
&\\
20200301 & Hospital in Gorinchem in quarantine\\
20200304 & Groningen students return from holiday in infected area\\
20200305 & Health minister Bruins informs parliament on COVID-19\\
20200306 & First COVID-19 death in The Netherlands\\
20200310 & Dutch prime minister asks everyone to stop shaking hands\\
20200311 & WHO declares COVID-19 to be a pandemic\\
20200312 & Schools remain open after announcing first national measures\\
20200313 & Hoarding in Dutch supermarkets\\
20200314 & People stay at home\\
20200315 & New measures: schools, bars and restaurant close\\
20200316 & Prime-minister addresses the nation\\
20200317 & Applauding event for health care workers\\
20200318 & RIVM boss Van Dissel speaks with parliament\\
20200319 & Calls for more intensive car beds\\
20200320 & King addresses the nation\\
20200321 & Calls for social distancing\\
20200322 & Calls for staying at home\\
20200323 & Tougher measures to battle the pandemic\\
20200325 & Virus detected in Dutch sewer\\
20200326 & Problems in supply of test fluid from Swiss company\\
20200327 & KLM resumes flights to COVID-19-hit countries\\
20200330 & EU provides financial support to Morocco to fight COVID-19\\
20200331 & Prolongation of COVID-19 measures\\
\end{tabular}

\vfill
\pagebreak

\begin{tabular}{lp{10cm}}
20200401 & Debate in the parliament about COVID-19\\
20200403 & RIVM rejects testing on Schiphol\\
20200405 & Germany declares Netherlands as risk area\\
20200407 & National press conference announces work on COVID-19 app\\
20200410 & Promising Israeli treatment\\
20200411 & Netherlands sent face masks to China in February\\
20200414 & Call for extra protection for health care workers\\
20200416 & Disturbances of youths in Monnickendam\\
20200419 & COVID-19 app proves to be unsafe\\
20200420 & Denmark put restrictions on companies asking for support\\
20200421 & National press conference\\
20200423 & Netherlands supplied face masks to Montenegro\\
20200424 & Trump recommends disinfectant as medicine\\
20200425 & China removes critical notes from WHO reports\\
20200426 & Nurse complains about lack of public compliance to measures\\
20200427 & Kings Day: calls to stay at home\\
20200428 & Complaints about number of people allowed in IKEA shops\\
20200429 & Public offers to order face masks for government\\
20200430 & Schiedam storage contains face masks to be send abroad\\
&\\
20200501 & Marseille professor cures people with hydroxychloroquine\\
20200502 & Dutch company DSM is involved in producing face masks\\
20200504 & Different celebration of WWII death remembrance\\
20200505 & Different celebration of Liberation Day\\
20200506 & National press conference: face masks in public transport\\
20200508 & Incident at Rotterdam supermarket about COVID-19 measures\\
20200509 & RIVM disapproved testing in care homes\\
20200510 & Warning for illness effects on young people\\
20200511 & Relaxation of COVID-19 measures\\
20200512 & RIVM refuses to publish reproduction number\\
20200514 & Belgian economist proposes COVID-19 tax for elderly\\
20200515 & Government party VVD profits from COVID-19 strategy in polls\\
20200516 & Government bought bad face masks\\
20200517 & RIVM has movie with medicine claims removed from YouTube\\
20200518 & Dutch doctor claims success of anti-malaria medicine\\
20200519 & National press conference\\
20200520 & COVID-19 debate in parliament\\
20200521 & Calls to stay at home, despite the weather\\
20200522 & Government thinks about postponing 2021 elections\\
20200523 & Church service contamination in Frankfurt\\
20200524 & Different celebration of Ramadan ending\\
20200525 & Unknown children's disease, possible COVID-19 link\\
20200526 & Discussion about aerosol contamination, influence of ventilation\\
20200528 & Netherlands has more per capita COVID-19 deaths than the USA\\
20200529 & Government works on emergency law on phone location data research\\
20200530 & COVID-19 outbreak in The Hague mosque\\
20200531 & Preparing for Black Lives Matter demonstration in Amsterdam\\
\end{tabular}

\vfill
\pagebreak

\begin{tabular}{lp{10cm}}
20200601 & Black Lives Matter demonstration in Amsterdam\\
20200602 & Call to cancel all COVID-19 fines\\
20200603 & Black Lives Matter demonstration in Rotterdam\\
20200604 & COVID-19 debate in parliament\\
20200605 & Stand-up comedian Hans Teeuwen includes COVID-19 in show\\
20200606 & Government discusses COVID-19 law\\
20200607 & Patient reports second contamination\\
20200609 & Critique on emergency COVID-19 law\\
20200610 & Dutch vaccine enters testing phase in July\\
20200611 & Protest of fancy fair workers\\
20200612 & COVID-19 causes long-term lung problems\\
20200614 & Volkskrant editor says that science, government and media must tell same story\\
20200616 & Mayors publish manifest on effects COVID-19 on society\\
20200617 & RIVM announces that finding contamination sources is hard\\
20200619 & The Hague forbids anti-COVID-19 measures demonstration\\
20200620 & NRC newspaper publishes timeline of Dutch pandemic experiences\\
20200621 & The Hague demonstration against COVID-19 measures still held\\
20200622 & RIVM forbids ventilator usage in health care homes\\
20200623 & Parliament forbids breading animals sensitive for COVID-19\\
20200624 & National press conference\\
20200625 & COVID-19 debate in parliament\\
20200627 & Call to stop social distancing rule\\
20200628 & Poll reveals disapproval (87\%) of governments COVID-19 policies\\
20200629 & Dutch Railways cuts 2,300 jobs\\
20200630 & Government parties vote against increasing health care worker salaries\\
&\\
20200702 & Doctors demand motivations for national COVID-19 measures\\
20200705 & Nurse Boy Eddema dies from COVID-19\\
20200706 & People demonstrate against incorrect COVID-19 news\\
20200708 & National COVID-19 app is named: CoronaMelder\\
20200709 & Medical staff protests against national pandemic measures\\
20200710 & Dordrecht forbids demonstration against COVID-19 law\\
20200711 & Health minister forced doctor to give satisfying intensive care bed estimate\\
20200712 & French bus driver killed in incident about face masks\\
20200713 & 23rd Dutch mink farm with COVID-19 found\\
20200714 & Mob profited from COVID-19 relief fund for companies\\
20200715 & Calls for testing even with mild symptoms\\
20200717 & 38 Schiphol-bound flights contained COVID-19 patients\\
20200719 & No passenger checks in Schiphol flights from contaminated areas\\
20200722 & Calls for more face mask wearing\\
20200723 & Discussion about acheivability of face mask obligation\\
20200725 & Despite reports, three-year old did not die of COVID-19\\
20200726 & Doctor protests against national COVID-19 measures\\
20200727 & Scar tissue found on hearts of COVID-19 patients\\
20200728 & Calls for keeping following the COVID-19 measures\\
20200729 & CNS reports that excess deaths is double of COVID-19 deaths\\
20200730 & Amsterdam and Rotterdam ask people in busy areas to wear face masks\\ 
\end{tabular}

\begin{tabular}{lp{10cm}}
20200801 & demonstration in Berlin against COVID-19 measures\\
20200813 & coalition leaves parliament to avoid vote on health care salaries\\
20200814 & police closes Twente supermarket for breaches of measures\\
20200815 & COVID-19 blogger Maurice de Hond banned from LinkedIn\\
20200824 & King criticized for lack of social distancing\\
20200827 & minister Grapperhaus criticised for lack of social distancing\\
&\\
20200901 & national press conference\\
20200909 & fires in Greek refugee camp Moria\\
20200915 & yearly speech by the King (Prinsjesdag)\\
20200921 & Dutch VIPs join protest against measures (\#ikdoenietmeermee)\\
20200922 & Dutch VIPs leave protest against measures\\
20200924 & social distancing fines dropped from criminal record\\
20200925 & hydroxychloroquine approved as anti-covid medicine\\
20200928 & national press conference\\
& \\
20201002 & Donald Trump tests positive for COVID\\
20201010 & national launch of the corona app CoronaMelder\\
20201013 & advice to wear face masks in shops\\
20201015 & King travels to Greece for a holiday\\
& \\
20201102 & several Ajax players test positive\\
20201103 & national press conference\\
20201104 & minks infect Danes with COVID\\
20201116 & Hungary and Poland block EU emergency fund\\
20201117 & national press conference\\
20201130 & start mandatory face mask wearing tomorrow\\
\end{tabular}

\end{document}